\documentclass{article}
\usepackage{spconf,amsmath,graphicx}
\usepackage{xcolor}

\usepackage{xspace}
\newcommand{\prjname}{HEiMDaL\xspace}

\title{\prjname: Highly Efficient Method for Detection and Localization of wake-words}
\name{Arnav Kundu, Mohammad Samragh Razlighi, Minsik Cho, Priyanka Padmanabhan, Devang Naik}
\address{\{a\_kundu, m\_samraghrazlighi, minsik, priyanka\_padmanabhan, naik.d\}@apple.com }
\begin{document}
%
\maketitle

\begin{abstract}
Streaming keyword spotting is a widely used solution for activating voice assistants. Deep Neural Networks with Hidden Markov Model (DNN-HMM) based methods have proven to be efficient and widely adopted in this space, primarily because of the ability to detect and identify the start and end of the wake-up word at low compute cost. However, such hybrid systems suffer from loss metric mismatch when the DNN and HMM are trained independently. Sequence discriminative training  cannot fully mitigate the loss-metric mismatch due to the inherent Markovian style of the operation. We propose an low footprint CNN model, called \prjname, to detect and localize keywords in streaming conditions. We introduce an alignment-based classification loss to detect the occurrence of the keyword along with an offset loss to predict the start of the keyword. \prjname shows 73\% reduction in detection metrics along with equivalent localization accuracy and with the same memory footprint as existing DNN-HMM style models for a given wake-word. 
\end{abstract}

\section{Introduction}

Voice assistants allow users to control electronic devices via vocal commands. In this setting, a device waits for the user to say a wake-word, e.g., ``hey Siri/ Alexa'', which indicates the user's intention to engage with a voice assistant. Then, the device records the remainder of the user's utterance and transmits it to a (possibly remote) voice assistant.  Since a wake-word recognizer often runs continuously on a device with small SRAM storage, the recognizer has to be parameter-efficient (i.e., it should have high model accuracy with few weights). Additionally, to respect users privacy, the recognizer should be accurate: it should only start recording the user's voice when the user intends to interact with the voice assistant. 

Several contemporary wake-word detection systems utilize a Deep Neural Network (DNN) together with a Hidden Markov Model (HMM)~\cite{sigtia2018efficient,chen2014small, chen2013hybrid}. In this setting, the DNN component is trained to identify word fragments (a.k.a. phonemes) while the HMM component traverses the sequence of phonemes predicted by the DNN and detects the wake-word. The combination of a DNN and an HMM can detect wake-words and their exact occurring time. However, such a hybrid system may suffer from loss metric mismatch: the DNN component is trained to detect the phoneme sequence, not the keyword itself, hence, the trained model may be sub-optimal. Sequence discriminative training is proposed in \cite{shrivastava2021optimize} where the DNN-HMM model is optimized end to end to minimize (maximize) the final HMM score for negative (positive) samples respectively. However, such models are difficult to optimize because of gradient loss during back propagation through the HMM. In addition, 
training DNN-HMM models takes substantially longer
due to the sequence dependent nature. More recent efforts train end-to-end CNNs to detect the underlying wake-word without an HMM~\cite{higuchi2020stacked, alvarez2019end}. These models can yield a good performance in wake-word detection as they are directly optimized to detect the wake-word but such CNNs suffer with two major limitations: \textbf{a)} higher computational complexity than DNN-HMM based systems, \textbf{b)} cannot accurately locate the exact occurrence time of the keyword.

We introduce \prjname, a wake-word detection system that simultaneously inherits the benefits of DNN-HMMs and end-to-end models yet with extremely low memory footprint. In particular, we train an end-to-end model that: \textbf{a)} does not utilize an HMM, \textbf{b)} is directly trained to detect the wake-word, \textbf{c)} is capable of predicting the start and end-time of the wake-word, and \textbf{d)} is more efficient and accurate than existing systems. To this end, we make the following contributions:
\begin{itemize}
    \item We formulate a discriminative setting for training an end-to-end wake-word detection model. We train the DNN to predict a binary label for a given segment of audio (the receptive field of the network) and an offset label to predict the start of the wake-word. 
   \item We introduce a localization-enforced classification loss along with a data mining algorithm. During training, we minimize our customized loss over the samples drawn by the mining algorithm. Our mining algorithm balances the positive and negative samples.  
    \item Compared to a DNN-HMM model \cite{shrivastava2021optimize}, our model improves the False Reject Rate by 73\% at the same False Accept Rate and model size in ``hey Siri'' detection.
\end{itemize}

\section{Related Work}
DNN-HMM models perform wake-word detection in two steps: the DNN converts the user's voice into the a sequence of phoneme probabilities. Next, the HMM traverses the phoneme probability sequence to detect the occurrence of a wake-word. In early DNN-HMM models, the DNN component is trained  to predict classification labels for the wake-word's phonemes, then  an HMM is applied at the inference phase to detect the wake-word~\cite{sigtia2018efficient,chen2014small, chen2013hybrid}. Other derivatives of this approach utilize alternative decoders in lieu of the HMM~\cite{prabhavalkar2015automatic, gruenstein2017cascade}. More recently, ~\cite{shrivastava2021optimize} noted that there is a mismatch between the training objective (phoneme classification) and the inference objective (wake-word detection), thus the phonemes detected by the DNN become more useful in detecting the wake-word when the HMM model is also plugged into the training loop. The   observation in~\cite{shrivastava2021optimize} has inspired us to move the discriminative training to, \prjname: we train a single DNN model which, unlike the DNN-HMM system, does not predict phoneme sequences. Our system is optimized to directly detect the wake-word and, as we show in our experiments, is more accurate than the DNN-HMM pipeline.

Recurrent neural networks~\cite{fernandez2007application,lengerich2016end,hwang2015online} have been utilized for wake-word detection but their computational complexity is often high. End-to-end CNN-based models are trained to detect the keyword directly~\cite{zhang2017hello, tang2018deep, sainath2015convolutional, alvarez2019end, higuchi2020stacked}. The aforementioned papers represent a few of many CNN-based methods in the literature, each of which may have been trained on different proprietary or public datasets. An important property of an end-to-end CNN-based wake-word detector is its compatibility with an audio streaming scenario, in which the input audio can have an arbitrary length. Among the existing CNN-based papers, we find~\cite{alvarez2019end} relevant to our streaming scenario: a singular value decomposition filter (SVDF) is introduced as an efficient layer for streaming wake-word detection, and  the CNN model is trained to detect the end frames of the wake-word by hard labeling. Followup work by~\cite{higuchi2020stacked} extends their architecture and evaluates the model for wake-word detection. The success of~\cite{higuchi2020stacked,alvarez2019end} motivated us to invest more in discriminative training for streaming wake-word detection. Compared to these works, our model can achieve a better detection accuracy, which can be attributed to our novel labeling approach, loss-function definition, and data sampling method during training. In addition, our system is capable of both detection and accurate localization of the wake-word, which is critical to ensure user privacy.

\section{Model Architecture}
\begin{figure}
    \centering
    \includegraphics[width=8.4cm]{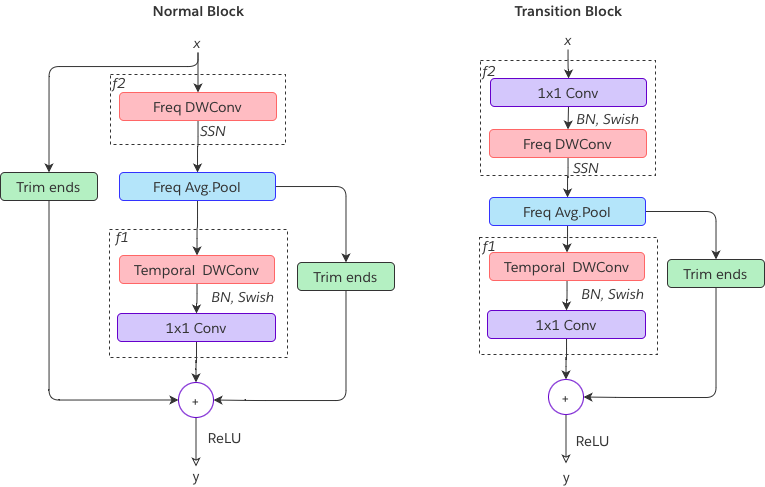}
    \caption{Modified BC-ResNet blocks}
    \label{fig: architecture}
\end{figure}

In \prjname experiments, we use  a modified version of BC-ResNet \cite{kim2021broadcasted} shown in Figure \ref{fig: architecture} for discriminative training. For an audio segment equal to the model's receptive field it produces two outputs: \textbf{a)} the probability that the segment ends with the wake-word,  \textbf{b)} the relative distance of the start of the wake-word from the end of the receptive field. We do not apply any padding in the depth-wise convolution layers and trim the skip connections from both ends such that output dimensions of the skip connections match to those of the convolutional layers of the block.

The overall model architecture is described in Table \ref{table:model}. Ignoring the batch dimension, at training time, the model consumes an input $x \in R^{(1\times C \times T)}$ and produces an output $z \in I^{(2\times 1 \times 1)}$; where $C \text{ and } T$ represents the frequency and time axis respectively. At inference time, the model can receive an input with arbitrary length $x \in R^{(1\times C \times (T+L))}$ and generate output $z \in I^{(2\times 1 \times L)}$. As such, the model can process a streaming audio at inference.
\begin{table}
\begin{tabular}{|c|c|c|c|c|c|}
\hline
     Input & Layer & n & c & s & d \\
     \hline
     $1 \times 16 \times 131$& conv2d 5x5 & 1 & 12 & 2,1 & 1,1 \\
     $12 \times 8 \times 127$& Transition Block & 1 & 16 & 1,1 & 1,1\\
     $16 \times 8 \times 125$& Broadcasted Block & 1 & 16 & 1,1 & 1,1\\
     $16 \times 4 \times 123$& Transition Block & 1 & 16 & 2,1 & 1,2\\
     $16 \times 4 \times 119$& Broadcasted Block & 1 & 16 & 1,1 & 1,2\\
     $16 \times 4 \times 115$& Transition Block & 1 & 32 & 1,1 & 1,4\\
     $32 \times 4 \times 107$& Broadcasted Block & 3 & 32 & 1,1 & 1,4\\
     $32 \times 4 \times 91$& Transition Block & 1 & 16 & 1,1 & 1,8\\
     $16 \times 4 \times 83$& Broadcasted Block & 3 & 16 & 1,1 & 1,8\\
     $16 \times 4 \times 19$& Transition Block & 1 & 16 & 1,1 & 1,4\\
     $16 \times 4 \times 11$& Broadcasted Block & 1 & 16 & 1,1 & 1,4\\
     $16 \times 4 \times 3$& conv2d 3x3 & 1 & 16 & 1,1 & 1,1 \\
     $16 \times 2 \times 1$& conv2d 1x1 & 1 & 8 & 1,1 & 1,1 \\
     $8 \times 2 \times 1$& conv2d 2x1 & 2 & 1 & 1,1 & 1,1 \\
     \hline
\end{tabular}
\caption{BC-ResNet: Each row describes a type of layer used, its corresponding input dimension, number of repeated blocks $n$, number of output channels $c$, stride ($s$) and dilation ($d$) used for the frequency or temporal convolution used in them.}
\label{table:model}
\end{table}
\section{Mining positive and negative samples}
\begin{figure}
    \centering
    \includegraphics[width=9cm]{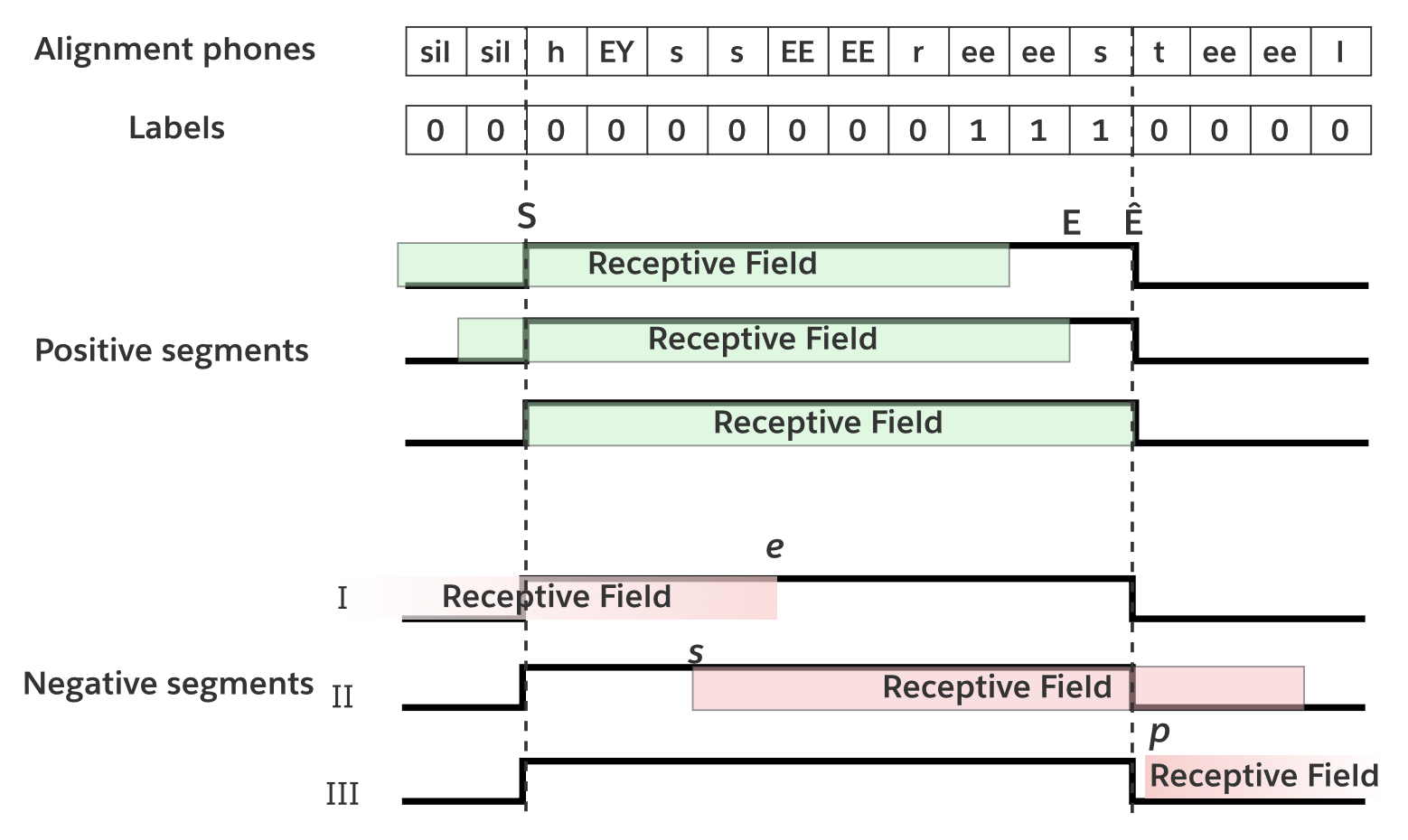}
    \caption{Mining positive and negative segments from given phonetic alignments}
    \label{fig:datamining}
\end{figure}

Each training utterance is forced aligned to find the per-frame phonetic alignments using an acoustic model. This enables finding the start ($S$) and end ($E$) of the wake-word during creation of training samples as shown in Figure \ref{fig:datamining}. Given the start and end of the wake-word, we label all those frames which match the last phone of the wake-word as 1. The same phone is labelled as 0 when it appears in a context other than the wake-word. For example, the frames with label $ee$ are marked as 1 when in context of the wake-word $h$ $EY$ $s$ $EE$ $r$ $ee$ but as 0 when in context $s$ $t$ $ee$ $ee$ $l$.  To account for alignment errors we mark a few frames after the end of wake-word as 1 depending on the number of repetitions of the last phone before $E$. The new end is marked as $\hat{E}$. 

The labels we discuss up to here are per-frame labels, which are also used by prior work~\cite{higuchi2020stacked, alvarez2019end}. We note here that each output of a convolutional network corresponds to an audio segment of length $T$, not a single frame. Hence, the training labels should not be generated solely based on single frames. Rather, the labels should also be assigned based on the overlap between the ground-truth wake-word and the underlying audio segment. Therefore, instead of relying on per-frame labels, we define per-segment labels. We define a positive example as an audio segment of length $T$ such that:
\begin{itemize}
    \item The entire wake-word falls in the segment 
    \item The segment ends with a frame labelled as 1
\end{itemize}
We extract positive segments with the above criteria and use them during training. Since we are training a discriminative model, we also need to create strong negative data so that the model doesn't learn to classify any word ending with the last few phones of the wake-word as a positive sample. We mine 3 types of negative samples in which:
\begin{enumerate}
    \item Segments end before end of the wake-word \label{item:segment1}
    \item Segments end after end of the wake-word \label{item:segment2}
    \item Segments start and end after end of the wake-word \label{item:segment3}
\end{enumerate}
Figure \ref{fig:datamining} illustrates the method adopted for mining positive segments and three types of negative segments. The end ($e$) of the Type \ref{item:segment1} segments is selected by picking random indices between start of the wake-word and the last but one phonetic index of the wake-word. Then we pick $X \in x(e-T,e)$ as the input to our model where $T$ is the receptive field of the model. Similarly, we select the start ($s$) of Type \ref{item:segment2} segments by picking random indices between the second phone of the wake-word and end of the wake-word. The resulting input segment is $X \in x(s,s+T)$.  For Type \ref{item:segment3} segments we select $X \in x(p,p+T)$ such that $p>\hat{E}$ where $\hat{E}$ denotes the end of the wake-word. If the length of a segment is lesser than the receptive field as denoted by the shaded blocks in Type \ref{item:segment1} and \ref{item:segment3} in figure \ref{fig:datamining} we pad the segments with random noise or silence.
\section{Results and Observations}
We apply our method for ``hey Siri'' detection. The remainder if this section explains the dataset, training, and comparison of \prjname with existing methods.

\subsection{Datasets}
Our training data consist of 500k utterances each containing the wake-word along with a user query that follows the wake-word. These audio samples are augmented using room-impulse responses (RIRs) and echo residuals. We also use ambient noise samples from various acoustic environments to augment the training data. We also apply training time gain augmentation of 10dB to -40dB during training. The positive test data contain both near and far-field utterances (captured at 3ft and 6ft distance) with ``hey Siri'' in them. The negative test data contain around 2000 hours of dense speech. These datasets have been collected through internal user studies with informed consent approvals.
\subsection{Training and Evaluation}
Given an input segment, the model is trained for two tasks:
\begin{itemize}
    \item Classification: Probability that the segment contains the wake-word at the end.
    \item Regression: Distance of start of wake-word from the segment end.
\end{itemize}
For the classification task, we use focal loss~\cite{lin2017focal} especially because of the skewed distribution of positive and negative samples. For the regression task we use mean squared error loss. The overall loss function is defined in Equation \ref{eq:loss}
\begin{equation}
    \label{eq:loss}
    L=(1-\sigma(\hat{y}))^{\gamma}\times BCE(y,\hat{y}) \\
    + MSE(d,\hat{d})\times I(y=1)
\end{equation}
where $\hat{y}$ is the ground-truth label, $y$ is the wake-word probability predicted by the network, $\sigma$ is the sigmoid function, $\gamma$ is a focal-loss hyper-parameter (4 in our case), BCE denotes cross-entropy, MSE denotes mean-squared error, and $I(y=1)$ is a binary filter that is $1$ where the label is positive. We optimize the proposed loss in Eq.~(\ref{eq:loss}) using Adam optimizer \cite{kingma2014adam} with learning rate of 0.01 with a cosine-annealing scheduler for 100 epochs. We use data distributed training 16 GPUs (2nodes x 8 GPU configuration) with a per GPU batch size of 64 utterances. Using our data sampling technique we select 1 positive  segment and 20 negative segments from each utterance making the effective batch size per GPU as 1344. Our training being non sequential takes $\approx 50\%$ less time than \cite{shrivastava2021optimize}. The input audio samples are augmented using gain and ambient noise augmentation on the fly in time domain. These samples are then transformed into 16 MFCC components for every 250 ms of audio at a 100 ms frame rate.

\begin{figure} [!t]
\vspace{-0.25in}
    \centering
    \includegraphics[width=9.3cm]{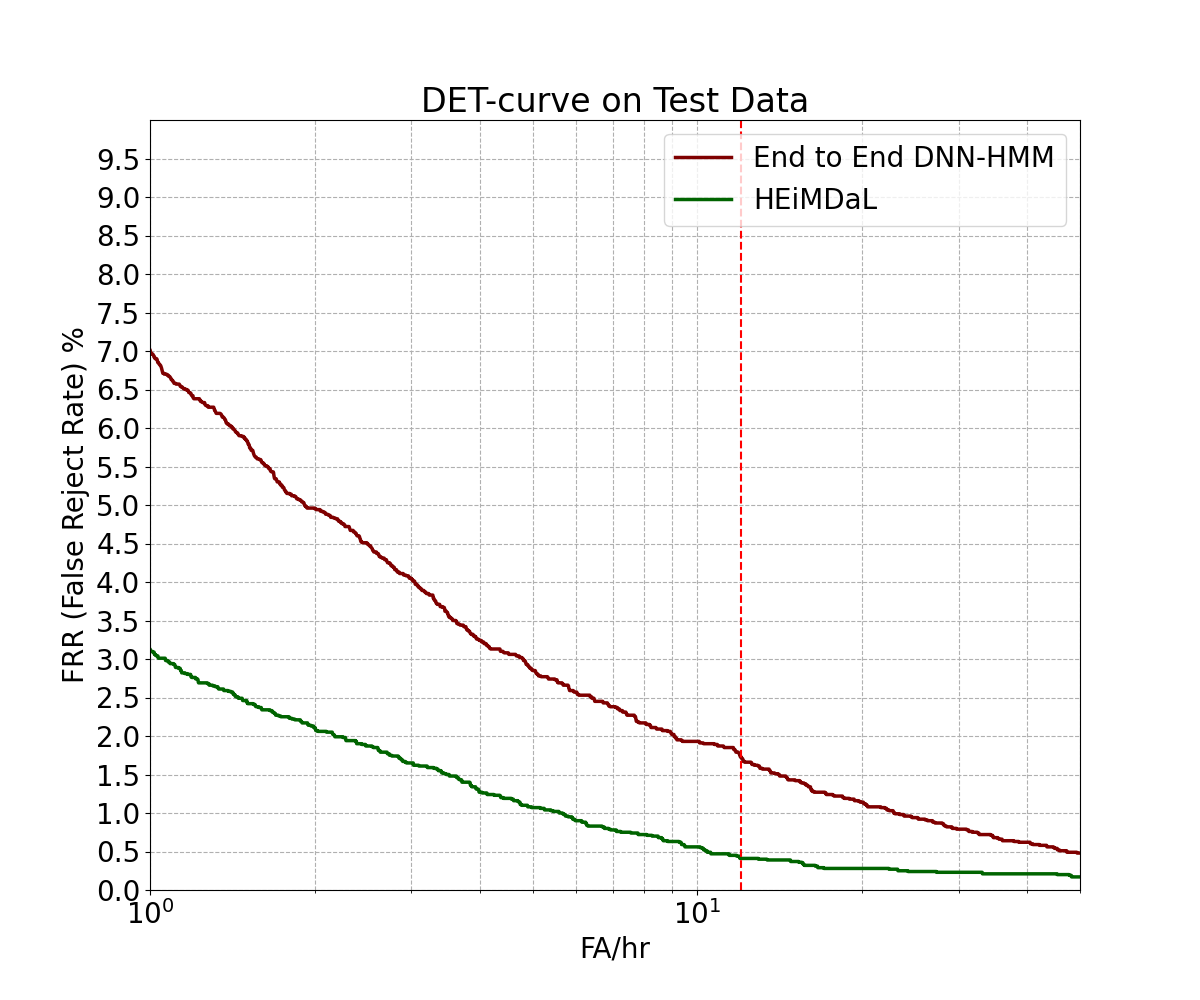}
    \caption{Comparison of DET-curves obtained by End to End trained DNN-HMM model and \prjname.}
    \label{fig:detcurve}
\end{figure}

During evaluation, we consider a detected keyword as a true positive (TP) if it overlaps with the ground-truth window, otherwise we count it as a false accept (FA). All positive windows where the detector does not trigger are counted as false rejects (FRs). The detection threshold is varied to compute the detection error trade-off (DET) curve (FRR vs FA/hr). Figure~\ref{fig:detcurve} compares \prjname to End-to-end trained DNN-HMM \cite{shrivastava2021optimize}. As seen, our method outperforms the prior work by a large margin at different operating points. We also report the FRRs of all the proposed models at 12FA/hr operating point in Table \ref{tab:results} as compared to S1-DCNN \cite{higuchi2020stacked} and End-to-end trained DNN-HMM \cite{shrivastava2021optimize}. Compared to the best of the two prior works, our method reduces the FRR from $1.7\%$ to $0.45\%$, which yields about $73\%$ relative FRR improvement.

\begin{table}[!t]
    \centering
    \begin{tabular}{|c|c|c|}
    \hline
         Method & \#params & FRR\%  \\
         \hline
         S1DCNN \cite{higuchi2020stacked} & 13993 & 1.99\\
         End-to-end DNN-HMM \cite{shrivastava2021optimize} & 13979 & 1.7\\
         \prjname (\textbf{Ours}) & 13832 & 0.45\\
         \hline
    \end{tabular}
    \caption{Comparison of FRRs at 12 FA/hr on test data. Our model is 70\% better than the previous best model.}
    \label{tab:results}
\end{table}

\begin{figure}[!t]
    \centering
    \includegraphics[width=9.3cm]{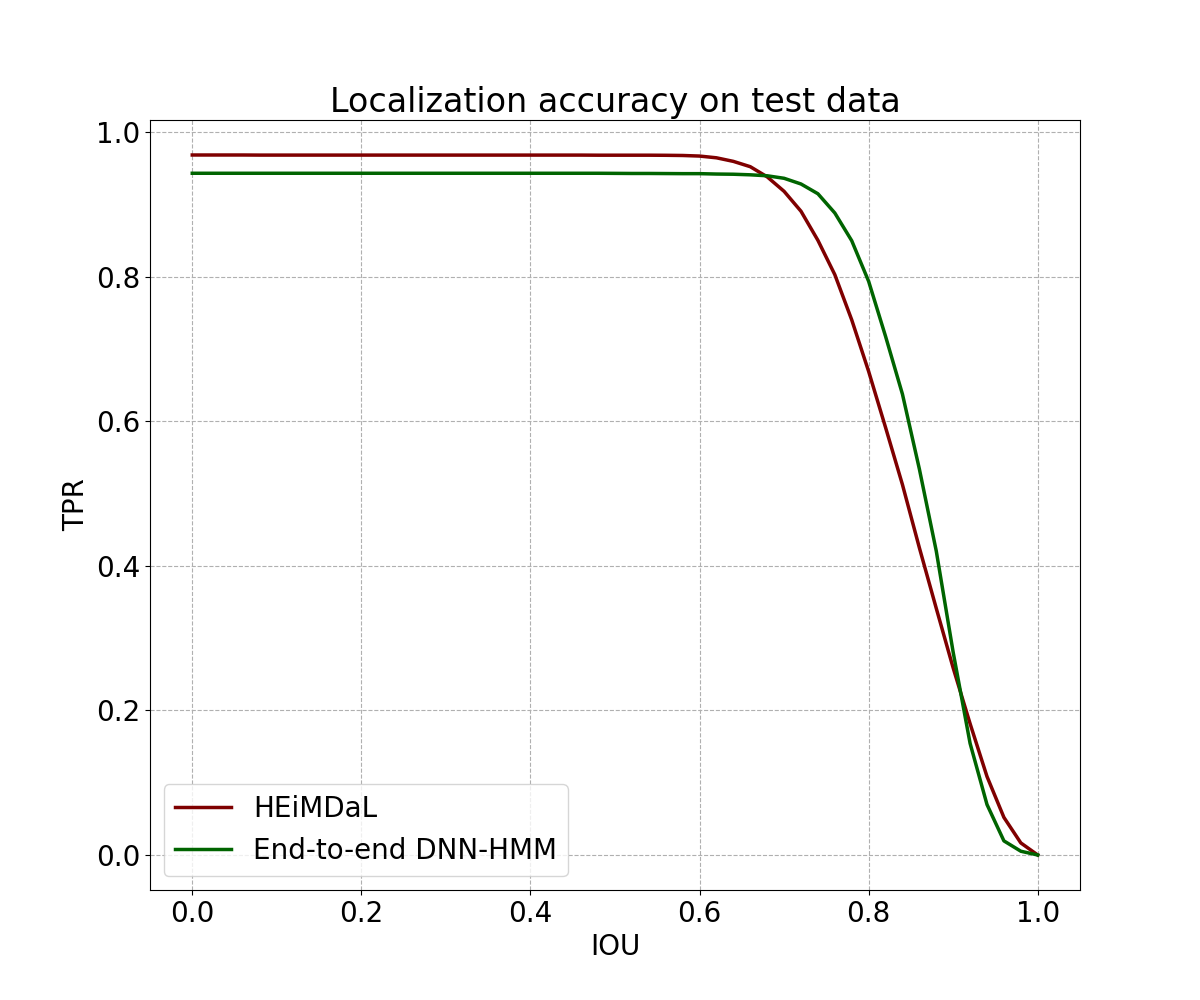}
    \caption{Comparing IOU vs TPR of End to End trained DNN-HMM model and \prjname.}
    \label{fig:localization}
\end{figure}

We also introduce a localization metric by plotting Intersection Over Union (IOU) vs True Positive Rate (TPR) for a baseline End to End trained DNN-HMM model and our \prjname model. The trigger end is identified by the point where the score from our models exceed the threshold at 12 FA/hr operating point for them. \prjname uses the offset predicted at that point to find the start of the the wake-word whereas the DNN-HMM model uses the trace of the HMM to find the start. Figure \ref{fig:localization} demonstrates the localization accuracy of both the methods where the Area Under Curve (AUC) for End to End DNN-HMM and \prjname is around 0.8.

\section{Conclusion}
In this work, we have introduced \prjname, a discriminative model for detection and localization of a wake-word in streaming speech by exhaustive data sampling and a localization enforcing discriminative loss. The proposed method works significantly better (73\% relative reduction in FRR) than a sequence discriminatively trained model while also reducing the training time by 50\%. We also showed that the localization performance of \prjname is at par with End to end DNN-HMM models.



\bibliographystyle{IEEEbib}
\bibliography{refs}

\end{document}